\documentclass[11pt, letterpaper]{article}
\usepackage[left=1in, right=1in, top=1in, bottom=1in]{geometry}
\usepackage[utf8]{inputenc}
\usepackage[T1]{fontenc}
\usepackage{bm}
\usepackage{type1cm}
\usepackage{lettrine}
\usepackage{amsmath,amssymb,amsthm}
\usepackage{moreverb}
\usepackage{mathtools}
\usepackage{amsmath}
\usepackage{amssymb}
\usepackage{algorithmic}
\usepackage{graphics}
\usepackage{graphicx}
\usepackage{caption}
\usepackage{extarrows}
\usepackage{color}
\usepackage{framed}
\usepackage{wrapfig}
\usepackage{bm}
\usepackage{mathrsfs}
\usepackage{mathabx}
\usepackage{multirow}
\usepackage{longtable}
\usepackage{hyperref}
\usepackage{paralist}
\usepackage{indentfirst}
\usepackage{relsize}
\usepackage{extarrows}
\usepackage{lineno}
\usepackage{upgreek}
\usepackage{bm}
\usepackage{mwe}
\usepackage[dvipsnames,table]{xcolor}
\usepackage{booktabs}
\usepackage{authblk}
\usepackage{subcaption}
\usepackage{lettrine}
\usepackage{type1cm}
\usepackage{threeparttable}
\usepackage[sort&compress,numbers]{natbib}
\usepackage[figurename=Figure]{caption}
\usepackage[normalem]{ulem}
\usepackage{tcolorbox}
\usepackage{pgfplots}
\usepackage{arydshln} 
\usepackage{enumitem}
\usepackage{pifont}
\definecolor{yesgreen}{RGB}{0,140,0}
\definecolor{nored}{RGB}{200,0,0}
\usepackage{siunitx}
\usepackage{makecell}
\usepackage{placeins}    
\usepackage{float}       

\graphicspath{ {./Figure/} }
\usepackage[font=footnotesize,labelfont=bf]{caption}

\providecommand{\keywords}[1]{\textbf{\textit{Keywords: }} #1}

\hypersetup{
bookmarks=true,
bookmarksopen=true,
bookmarksnumbered=true,
unicode=false,
pdftoolbar=true,
pdfmenubar=true,
pdffitwindow=false,
pdfstartview={FitH},
pdftitle={My title},
pdfauthor={Author},
pdfsubject={Subject},
pdfcreator={Creator},
pdfproducer={Producer},
pdfkeywords={keywords},
pdfnewwindow=true,
colorlinks=true,
linkcolor=blue,
citecolor=blue,
filecolor=blue,
urlcolor=blue
}

\definecolor{natureDarkBlue}{RGB}{34, 84, 126}
\definecolor{natureMidBlue}{RGB}{68, 149, 212}
\definecolor{natureLightBlue}{RGB}{189, 215, 238}
\definecolor{natureGray}{RGB}{235, 235, 235}

\pgfplotsset{
    nature_style/.style={
        font=\sffamily\scriptsize,  
        width=6.0cm, height=5.5cm,  
        axis line style={draw=none}, 
        axis x line*=bottom,
        axis y line*=left,
        axis line style={black!80, line width=0.8pt},
        ymajorgrids=true,
        grid style={natureGray, dashed},
        tick align=outside,
        tick style={black!50},
        legend style={draw=none, fill=none}
    }
}

\definecolor{purple}{HTML}{c994c7}
\definecolor{navyblue}{RGB}{30,130,255}
\definecolor{citecolor}{RGB}{30,130,255}
\definecolor{lightgray}{gray}{0.9}
\definecolor{blanchedalmond}{rgb}{1.0, 0.92, 0.8}
\definecolor{cerise}{rgb}{0.871, 0.192, 0.388}

\definecolor{TaskBG}{HTML}{EFE6FF}        
\definecolor{StateBG}{HTML}{F5F5F7}       
\definecolor{ExpertBG}{HTML}{EAF7EA}      
\definecolor{IWMBG}{HTML}{FDECF3}         
\definecolor{SRBG}{HTML}{E6F2FF}          

\newtcolorbox{trainingexample}[2][]{%
  enhanced, breakable, colframe=black!12, colback=white, boxrule=2.5pt,
  arc=2pt, left=0pt, right=0pt, top=0pt, bottom=0pt,
  title={#2}, fonttitle=\bfseries, coltitle=black, #1}

\newcolumntype{L}[1]{>{\RaggedRight\arraybackslash}p{#1}} 
\newcolumntype{Y}{>{\RaggedRight\arraybackslash}X}        

\AtBeginDocument{%
  }

\definecolor{lemon}{HTML}{FDFFCC}

\definecolor{Gainsboro}{rgb}{0.86, 0.86, 0.86}

\definecolor{Gray}{gray}{0.95}
\definecolor{LightCyan}{rgb}{0.88,1,1}
\definecolor{dm-blue-500}{RGB}{0, 69, 177}
\definecolor{dm-purple-500}{RGB}{105,50,230}
\definecolor{dm-red-500}{RGB}{255,122,122}

\definecolor{backred}{RGB}{255, 190, 190}
\definecolor{backblue}{RGB}{220, 230, 250}

\newtcbox{\hlprimarytab}{on line, rounded corners, box align=base, colback=backblue, colframe=white, size=fbox, arc=3pt, before upper=\strut, top=-2pt, bottom=-4pt, left=-2pt, right=-2pt, boxrule=0pt}
\newtcbox{\hlsecondarytab}{on line, box align=base, colback=backred, colframe=white, size=fbox, arc=3pt, before upper=\strut, top=-2pt, bottom=-4pt, left=-2pt, right=-2pt, boxrule=0pt}

\newcommand{\headernodot}[1]{\noindent\textbf{#1}}
\newcommand{\header}[1]{\headernodot{#1.}}

\begin{document}
\title{\textbf{Analyzing and Correcting Benevolence Bias in Large Language Models}}

\author[1]{Yuanzi Li}
\author[1]{Junhao Wang}
\author[2]{Minghui Liu}
\author[1]{Boyi Li}
\author[1]{Bingchen Chen}
\author[1]{Zihang Tian}
\author[3]{\mbox{Jingyu Zhao}}
\author[4]{Yuhan Wang}
\author[1]{Lei Wang}
\author[1]{Pei Wang}
\author[1]{Jinchao Wu}
\author[1,$^*$]{Xu Chen}

\affil[1]{\small Gaoling School of Artificial Intelligence, Renmin University of China, Beijing, China}
\affil[2]{\small School of Computer Science and Engineering, Sun Yat-sen University, Guangzhou, China}
\affil[3]{\small School of Computer Science and Technology, Shandong University, Qingdao, China}
\affil[4]{\small School of Mathematical Sciences, Peking University, Beijing, China}
\affil[*]{Corresponding authors}

\date{}

\maketitle

\normalsize

\vspace{-18pt}
\begin{abstract}
Large language models (LLMs) are increasingly used as stand-ins for human respondents, from opinion polls and simulated survey participants to agent-based social simulations. These uses rest on one assumption: that conditioning a model on who a person is yields answers resembling those of real people from that group. Here we identify and measure \emph{benevolence bias}, a small but consistent tendency for aligned LLMs to lean toward the kinder, safer, more socially approved answer on value-laden survey questions. Across 18 widely used models, four social-science datasets (ANES, GSS, WVS, and a cross-cultural prospect-theory replication) and six psychological categories, we find that the bias is a stable model property, not a quirk of any one system: it points the same way across models, grows with model size, and traces to the post-training stage. Prompt language and framing change its size but never its direction, and a ``malicious persona'' stress test shows a one-sided limit: aligned models struggle to play people who are less kind, less prosocial or more harm-tolerant than average. The issue is thus not only a shifted average, but a narrowed range of people the model can imitate. The bias sits in the middle of the answer distribution rather than its tails, and survives changes in sampling temperature and simple prompted reflection. The encouraging news is that it is easy to diagnose and straightforward to fix: a light-touch contrastive calibration, which needs no retraining and works on black-box APIs, brings all six categories back to the human baseline. Our results give researchers a clear map of where aligned LLMs can already be trusted as human stand-ins, where they need care, and a ready-to-use method for closing the gap.
\end{abstract}


\vspace{12pt}
\section*{Introduction}
Large language models (LLMs) are increasingly used as stand-ins for human respondents across the social and behavioural sciences: to answer opinion polls, to simulate survey participants, to populate agent-based models of social processes, and to supply synthetic samples where collecting human data at scale is hard~\cite{aher2023using,argyle2023out,horton2023large,park2023generative}. These uses promise faster, cheaper and larger studies, and they rest on one idea: that asking a model to answer as a given kind of person yields answers resembling what real people of that kind would say. Whether today's aligned models live up to this idea, and on which questions, has never been tested systematically. Answering that question is what turns LLM-based simulation from a promising idea into a dependable method.

The question matters because the same models are also built for a different job: from chat assistants to policy tools, being ``helpful'' and ``harmless'' is a core design goal. Reinforcement learning from human feedback~\cite{christiano2017deep,ouyang2022training,stiennon2020learning}, supervised instruction tuning~\cite{wei2022finetuned}, and constitutional or preference-based safety training~\cite{bai2022training,bai2022constitutional,casper2023open,rafailov2023direct} all steer models toward answers that people find helpful, harmless and agreeable, rewarding prosocial wording, caution around possible harm, and widely acceptable choices~\cite{perez2023discovering,sharma2024towards}. This mark points in the same direction as the categories along which value-laden survey questions are scored~\cite{kirk2024benefits,wolf2024fundamental}. The training that makes models safer may therefore nudge the whole distribution of their attitudes, not just sharpen it. If so, any conclusion drawn from a synthetic sample inherits that nudge, and a researcher looking only at the model's answers has no way to notice. Seeing the nudge clearly and measuring it is the first step toward trusting these tools.

Earlier work has assembled pieces of this picture: today's models lean liberal on policy questions and favour certain demographic profiles~\cite{durmus2024towards,hartmann2023political,rozado2024political,santurkar2023whose}; fine-tuning changes behaviour on contested topics and value judgements~\cite{atari2023which,cao2023assessing,rottger2024political}; persona conditioning moves answers in the right direction but does not reliably reproduce the true distribution of a group~\cite{bisbee2024synthetic,gupta2024bias,sun2024random}; aligned models drift toward answers that match what the user seems to want~\cite{perez2023discovering,sharma2024towards}; and theory suggests alignment objectives cannot fully remove every unwanted behaviour~\cite{wolf2024fundamental}. What is still missing is a clear, category-by-category map of this alignment-driven shift: which value categories are affected; how it changes with model size and capability; whether prompt-level tools can move it; whether it belongs to the models or to the datasets; and whether it is only a shift in the average, or also a shrinking of the range of answers a model can produce. Without such a map, neither model builders nor downstream users can check for this bias or reason about it.

Two things make this map especially timely. First, exactly this generation of aligned models is being adopted at scale for synthetic samples and ``silicon sample'' experiments~\cite{argyle2023out,bisbee2024synthetic,sun2024random}, so any value shift is quietly passed along to a growing body of work. Second, established survey instruments such as the World Values Survey~\cite{inglehart2014wvs}, the General Social Survey~\cite{smith2018general}, the American National Election Studies~\cite{anes2020} and cross-cultural prospect-theory replications~\cite{ruggeri2020replicating} already provide a human reference, and they cover exactly the value domains alignment is most likely to touch. Together these supply both the motivation and the measuring stick.

\begin{figure*}[!tb]
  \centering
  \includegraphics[width=1\linewidth]{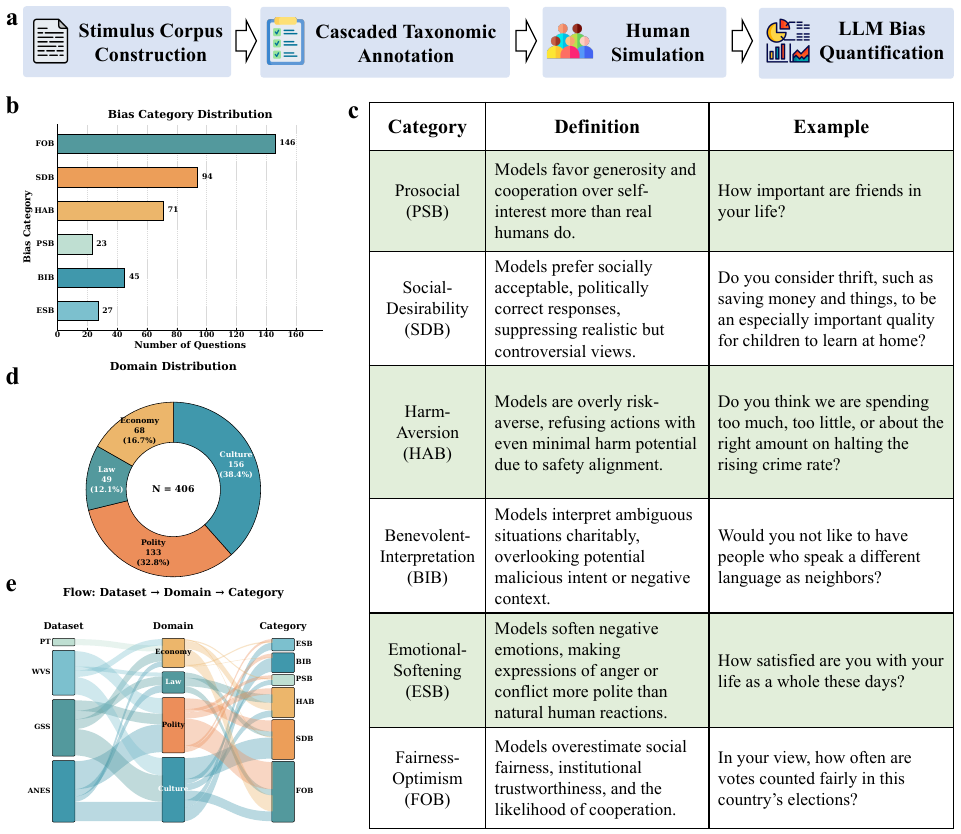}
  \caption{\textbf{Benchmark construction and taxonomy.}
  \textbf{a}, Overview of the benchmark pipeline. Value-laden survey items are collected from source corpora, assigned to a cascaded taxonomy of social domains and benevolence-bias categories, used to elicit simulated human responses from LLMs, and compared against empirical human response distributions to quantify the bias.
  \textbf{b}, Distribution of the 406 retained benchmark items across the six benevolence-bias categories: prosociality (PSB), social desirability (SDB), harm aversion (HAB), benevolent interpretation (BIB), emotional softening (ESB) and fairness optimism (FOB).
  \textbf{c}, Category definitions with representative survey items, illustrating the behavioural tendency captured by each bias dimension.
  \textbf{d}, Composition of the benchmark across the four social domains of culture, politics, law and economy, showing the number and proportion of items assigned to each domain.
  \textbf{e}, Alluvial flow from source dataset to domain and bias category, summarizing how ANES, GSS, WVS and PT items populate the final benchmark taxonomy.}
  \label{fig:benchmark}
\end{figure*}

Here we provide such a map across 18 widely used LLMs and four social-science datasets (ANES, GSS, WVS, and a cross-cultural prospect-theory replication; Fig.~\ref{fig:benchmark}a), with questions spanning four social domains and six benevolence-bias categories (Fig.~\ref{fig:benchmark}b,d,e). The six-category scheme splits ``benevolence'' into separate strands: social-desirability self-presentation~\cite{edwards1957social}, harm aversion, prosocial motivation, benevolent interpretation, fairness optimism, and emotional softening (Fig.~\ref{fig:benchmark}c). Two simple metrics track the bias: the Benevolence Tendency Bias (BTB), the overall shift at the population level, and the Benevolence Win Rate (BWR), the direction question by question. We then study the bias at three points in the pipeline: at the model level, whether it is stable, how it scales, and whether it traces to post-training; at the input level, whether it is just the default assistant voice or a deeper prior carried across languages, framings and personas; and at the output level, whether it lives only at the most likely answer or runs through the whole distribution, using contrastive calibration first as a measuring tool and then as a practical fix.

Three findings map the shape of the bias and how to handle it. First, it points consistently one way and lands hardest where alignment aims: 83 of 108 BTB cells are positive, 75 of 108 BWR cells sit above the baseline, all 18 models shift positively on social desirability and 17 of 18 on harm aversion; the bias grows with size in both Qwen families, is only weakly linked to capability ($R^2 = 0.16$), ranks models consistently across the three large surveys (Spearman's $\rho$ 0.54--0.63), and traces to post-training. Second, it is not just a shifted average: a ``malicious persona'' stress test can push models below their normal values on fairness optimism and emotional softening, but never below the human baseline on social desirability, prosociality or harm aversion; on those axes the models have lost some ability to imitate the less-benevolent end. Third, prompt-level tools change its size but never its direction, whereas contrastive calibration at a moderate setting pulls all six categories into a tight band around the human baseline. Alignment thus shifts, rather than merely polishes, the value distribution of today's models. That is encouraging news for practice: because the shift is predictable and distributional, a light-touch correction with no retraining can bring aligned LLMs back in line with the human reference, so they can be used responsibly as proxies for human respondents.

\section*{Results}
\subsection*{Model level}

\header{The bias is common but concentrated on a few dimensions} Across the four datasets (ANES, GSS, WVS and PT), we compared 18 models against the human baseline (BTB $=0$, BWR $=0.5$) on six bias dimensions (Table~\ref{tab:main}). Of the 108 BTB cells, 83 are positive; of the 108 BWR cells, 75 are above 0.5; the mean BTB is 0.027 and the mean BWR is 0.527. The bias thus points the same way across models and dimensions, though it is moderate in size. It is largest on social desirability, where every model is positive (mean BTB $=0.059$, mean BWR $=0.565$), and harm aversion, where 17 of 18 are (mean BTB $=0.066$, mean BWR $=0.569$). Prosociality, benevolent interpretation and fairness optimism are positive on average but smaller, and emotional softening shows no clear shift (mean BTB $=0.001$, mean BWR $=0.494$). Individual models still differ: GPT-4.1-mini, GPT-5-mini, Qwen3-32B, Qwen2.5-72B-Instruct and Llama-3.3-70B sit above the baseline on almost every dimension, whereas Gemini3-Flash, DeepSeek-V4-Pro, Mimo-v2.5 and Gemma-3-27B pair a strong social-desirability or harm-aversion shift with near-neutral or negative values elsewhere. Benevolence bias is thus not a uniform push along every value axis, but a positive shift focused on social approval, safety and harm avoidance.

\begin{figure*}[!tb]
  \centering
  \includegraphics[width=1\linewidth]{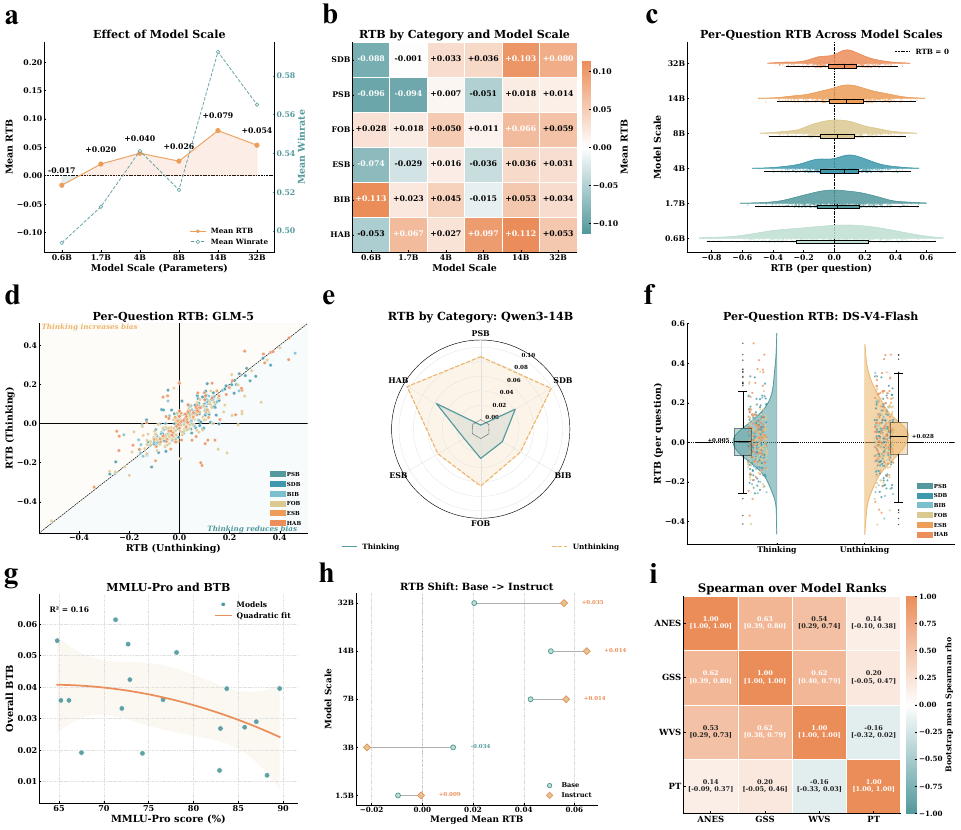}
  \caption{\textbf{Model-level patterns of benevolence bias.}
  \textbf{a}, Effect of model scale on the pooled mean BTB and mean win rate of Qwen models.
  \textbf{b}, Category-level BTB for the six bias dimensions across model scales.
  \textbf{c}, Per-question BTB distributions across model scales.
  \textbf{d}, Per-question BTB of GLM-5 under direct answering versus model reasoning, coloured by bias category.
  \textbf{e}, Category-level BTB profile of Qwen3-14B under direct answering versus model reasoning.
  \textbf{f}, Per-question BTB distributions of DeepSeek-V4-Flash under direct answering versus model reasoning.
  \textbf{g}, Relationship between MMLU-Pro and overall BTB, with a quadratic fit.
  \textbf{h}, Change in pooled mean BTB from base to instruction-tuned checkpoints for Qwen2.5 models at different scales.
  \textbf{i}, Bootstrap Spearman correlations of model rankings across datasets.}
  \label{fig:modellevel}
\end{figure*}

\header{The bias grows as models get larger} Within the Qwen3 family (0.6B, 1.7B, 4B, 8B, 14B and 32B), with datasets, prompting and scoring fixed, the mean BTB rises from close to the human baseline at 0.6B to a clearly higher value at 32B, and the mean win rate moves from about chance to well above 0.5 (Fig.~\ref{fig:modellevel}a). Larger models therefore give benevolent answers more often. The growth is uneven across categories (Fig.~\ref{fig:modellevel}b): social-desirability and fairness-optimism bias grow fastest, harm aversion turns from slightly negative in small models to positive in large ones, and the other categories change less. The per-question distributions show the same pattern (Fig.~\ref{fig:modellevel}c): as Qwen3 grows, the BTB distribution shifts right and develops a heavier positive tail, so the rising average reflects a growing share of high-bias questions rather than a uniform push. The Qwen2.5 family behaves similarly (Supplementary Information).

\header{Turning on model reasoning reduces the bias} We then tested whether letting a model ``think'' before answering changes the bias, using three families with a switchable thinking mode (GLM-5, Qwen3-14B and DeepSeek-V4-Flash) and holding questions, scoring and dataset mix fixed. Per-question BTB fell in all three models once thinking was on. For GLM-5, most points fall below the identity line, across social desirability, harm aversion, fairness optimism and benevolent interpretation (Fig.~\ref{fig:modellevel}d). The category profile of Qwen3-14B pulls in toward the neutral centre, with the biggest drops where the bias started strongest (Fig.~\ref{fig:modellevel}e). The per-question distribution of DeepSeek-V4-Flash shifts toward zero, with fewer strongly positive answers (Fig.~\ref{fig:modellevel}f). The improvement is partial: most categories keep a small positive residual, and no model fully matches the human reference. Reasoning softens the bias without removing it.

\header{More capable models show a slightly smaller bias} Relating each model's MMLU-Pro score to its overall BTB with a quadratic fit, the curve slopes gently downward: more capable models show smaller shifts on average, and the strongest systems come closest to the human baseline (Fig.~\ref{fig:modellevel}g). But the link is loose ($R^2 = 0.16$): models with similar scores can sit far apart on the BTB axis. Capability explains only a small part of the differences between models.

\header{Instruction tuning makes Qwen2.5 more benevolent} To pin down where in the pipeline the bias enters, we compared Qwen2.5 base models against their instruction-tuned versions at the same size, keeping the family and pre-training fixed (Fig.~\ref{fig:modellevel}h). Instruction tuning raised the mean BTB at the 1.5B, 7B, 14B and 32B sizes, with the biggest jump at 32B; the 3B model is the only exception, with a small negative change. The bias thus enters mainly at post-training, and the effect grows with model size.

\header{Model rankings agree best across the general survey datasets} To check that the rankings reflect a stable model property rather than any one dataset, we computed bootstrap Spearman correlations between model BTB rankings across the four datasets (Fig.~\ref{fig:modellevel}i). The three general survey datasets rank the models consistently: $\rho \approx 0.63$ for ANES--GSS, $\rho \approx 0.62$ for GSS--WVS, and $\rho \approx 0.54$ for ANES--WVS. Agreement is weaker for PT, which has fewer questions and a narrower range of topics, and correlates only weakly with the other three.

\begin{table}[!tb]
\centering
\caption{\textbf{Overall benevolence-bias results under the w/ inst.\ setting.} Each bias dimension is reported with Benevolence Tendency Bias (BTB) and Benevolence Win Rate (BWR). Positive BTB and BWR $>0.5$ indicate a displacement towards the benevolent pole relative to the human reference.}
\label{tab:main}
\footnotesize
\setlength{\tabcolsep}{2pt}
\begin{tabular}{l *{12}{S[table-format=-1.3]}}
\toprule
\multirow{2}{*}{Model} & \multicolumn{2}{c}{PSB} & \multicolumn{2}{c}{SDB} & \multicolumn{2}{c}{HAB} & \multicolumn{2}{c}{BIB} & \multicolumn{2}{c}{ESB} & \multicolumn{2}{c}{FOB} \\
\cmidrule(lr){2-3} \cmidrule(lr){4-5} \cmidrule(lr){6-7} \cmidrule(lr){8-9} \cmidrule(lr){10-11} \cmidrule(lr){12-13}
 & {BTB} & {BWR} & {BTB} & {BWR} & {BTB} & {BWR} & {BTB} & {BWR} & {BTB} & {BWR} & {BTB} & {BWR} \\
\midrule
\multicolumn{13}{l}{\textit{Closed-source models}} \\
GPT-5.1            &  0.026 & 0.531 & 0.087 & 0.600 &  0.077 & 0.582 & -0.002 & 0.500 & -0.029 & 0.447 & -0.014 & 0.473 \\
GPT-5-mini         &  0.057 & 0.558 & 0.072 & 0.578 &  0.059 & 0.556 &  0.004 & 0.510 &  0.025 & 0.510 &  0.019 & 0.516 \\
GPT-5.4-mini       & -0.015 & 0.474 & 0.064 & 0.576 &  0.067 & 0.574 &  0.001 & 0.502 &  0.020 & 0.516 &  0.022 & 0.521 \\
GPT-5.4-nano       & -0.069 & 0.382 & 0.062 & 0.571 &  0.026 & 0.515 & -0.009 & 0.484 &  0.025 & 0.523 &  0.006 & 0.496 \\
GPT-4o-mini        &  0.001 & 0.470 & 0.110 & 0.626 &  0.140 & 0.675 & -0.015 & 0.475 &  0.080 & 0.593 &  0.014 & 0.509 \\
GPT-4.1-mini       &  0.091 & 0.591 & 0.079 & 0.585 &  0.069 & 0.568 &  0.027 & 0.543 &  0.016 & 0.513 &  0.031 & 0.533 \\
Grok-4.1-Fast      &  0.002 & 0.504 & 0.037 & 0.543 & -0.009 & 0.490 &  0.016 & 0.529 & -0.007 & 0.483 &  0.029 & 0.529 \\
Gemini3-Flash      &  0.001 & 0.489 & 0.043 & 0.551 &  0.053 & 0.555 & -0.006 & 0.487 & -0.007 & 0.484 & -0.019 & 0.468 \\
Gemini2.5-Flash    &  0.030 & 0.538 & 0.049 & 0.563 &  0.068 & 0.575 & -0.002 & 0.495 & -0.034 & 0.455 & -0.026 & 0.466 \\
\addlinespace[2pt]
\multicolumn{13}{l}{\textit{Open-source models}} \\
Llama-3.3-70B      &  0.021 & 0.516 & 0.088 & 0.601 &  0.093 & 0.601 &  0.035 & 0.548 & -0.013 & 0.473 &  0.055 & 0.566 \\
Qwen3.7-Max        &  0.028 & 0.527 & 0.082 & 0.598 &  0.083 & 0.596 &  0.017 & 0.523 & -0.008 & 0.485 &  0.007 & 0.503 \\
DeepSeek-v4-Flash  & -0.010 & 0.480 & 0.029 & 0.530 &  0.078 & 0.585 & -0.020 & 0.481 &  0.031 & 0.534 &  0.020 & 0.518 \\
DeepSeek-V4-Pro    &  0.003 & 0.497 & 0.028 & 0.526 &  0.062 & 0.575 & -0.028 & 0.466 & -0.009 & 0.475 & -0.002 & 0.488 \\
GLM-5              &  0.021 & 0.507 & 0.052 & 0.552 &  0.060 & 0.555 & -0.006 & 0.490 &  0.006 & 0.496 &  0.009 & 0.501 \\
Kimi K2.6          &  0.012 & 0.499 & 0.047 & 0.542 &  0.057 & 0.561 &  0.005 & 0.504 & -0.009 & 0.482 &  0.010 & 0.504 \\
Mimo-v2.5-Pro      & -0.003 & 0.490 & 0.076 & 0.590 &  0.083 & 0.584 &  0.003 & 0.499 & -0.005 & 0.485 &  0.007 & 0.505 \\
Mimo-v2.5          &  0.005 & 0.495 & 0.051 & 0.558 &  0.087 & 0.588 & -0.005 & 0.490 & -0.011 & 0.469 & -0.013 & 0.474 \\
Minimax-M2.7       &  0.006 & 0.496 & 0.072 & 0.580 &  0.067 & 0.561 &  0.015 & 0.524 & -0.002 & 0.478 &  0.002 & 0.497 \\
Qwen3-32B          &  0.020 & 0.518 & 0.074 & 0.588 &  0.053 & 0.560 &  0.027 & 0.541 &  0.041 & 0.555 &  0.057 & 0.570 \\
Qwen3-235B-A22B    &  0.045 & 0.568 & 0.057 & 0.570 &  0.084 & 0.592 &  0.018 & 0.523 & -0.004 & 0.500 &  0.027 & 0.521 \\
Qwen2.5-72B-Inst   &  0.025 & 0.519 & 0.056 & 0.555 &  0.051 & 0.550 &  0.018 & 0.527 &  0.001 & 0.505 &  0.020 & 0.517 \\
Mistral-Small-24B  & -0.002 & 0.495 & 0.063 & 0.564 &  0.070 & 0.577 &  0.006 & 0.505 &  0.003 & 0.518 &  0.022 & 0.521 \\
Gemma-3-27B        & -0.050 & 0.440 & 0.052 & 0.555 &  0.085 & 0.586 &  0.010 & 0.504 & -0.027 & 0.455 & -0.014 & 0.477 \\
\midrule
Human              &  0.000 & 0.500 & 0.000 & 0.500 &  0.000 & 0.500 &  0.000 & 0.500 &  0.000 & 0.500 &  0.000 & 0.500 \\
\bottomrule
\end{tabular}
\end{table}

\subsection*{Input level}

\begin{figure*}[!tb]
  \centering
  \includegraphics[width=1\linewidth]{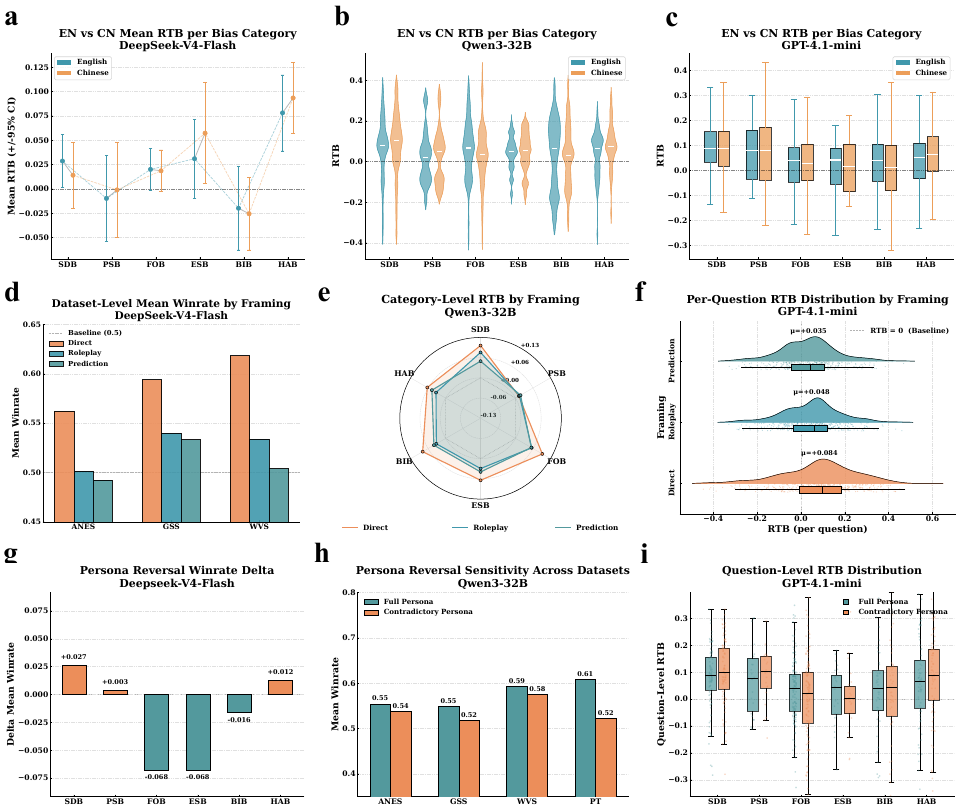}
  \caption{\textbf{Input-level modulators of benevolence bias.}
  \textbf{a}, English--Chinese comparison of mean BTB across bias categories for DeepSeek-V4-Flash, with 95\% confidence intervals.
  \textbf{b}, Category-level per-question BTB distributions of Qwen3-32B under English and Chinese prompts.
  \textbf{c}, Category-level per-question BTB distributions of GPT-4.1-mini under English and Chinese prompts.
  \textbf{d}, Dataset-level mean win rate of DeepSeek-V4-Flash under the direct-answering, role-play and prediction framings.
  \textbf{e}, Category-level BTB profile of Qwen3-32B under the direct-answering, role-play and prediction framings.
  \textbf{f}, Per-question BTB distributions of GPT-4.1-mini across prompt framings.
  \textbf{g}, Change in category-level mean win rate of DeepSeek-V4-Flash from the original to the malicious persona.
  \textbf{h}, Dataset-level mean win rate of Qwen3-32B under the original and malicious personas; the human baseline is 0.5.
  \textbf{i}, Question-level BTB distributions of GPT-4.1-mini under the original and malicious personas.}
  \label{fig:inputlevel}
\end{figure*}

\header{Prompt language changes the size of the bias, not its direction} To check whether the shift depends on the language of the prompt, we translated the four benchmarks into Chinese and re-ran DeepSeek-V4-Flash, Qwen3-32B and GPT-4.1-mini with matched scoring, decoding and question mix (Fig.~\ref{fig:inputlevel}a--c). For DeepSeek-V4-Flash, every dimension stays on the same side of the human baseline in both languages, and the English and Chinese 95\% confidence intervals overlap for five of six categories (Fig.~\ref{fig:inputlevel}a). For Qwen3-32B, harm aversion is higher under Chinese prompts and fairness optimism lower, though both stay positive (Fig.~\ref{fig:inputlevel}b). For GPT-4.1-mini, the two languages almost fully overlap (Fig.~\ref{fig:inputlevel}c). Switching language never flips the sign of the bias on any dimension; which dimensions show a language gap, and in which direction, differs by model. Prompt language is thus a dial, not a switch: each model's language-specific representations filter, but do not remove, the alignment-driven prior.

\header{Prompt framing changes the size of the bias without removing it} We next compared three framings with matched scoring, decoding and question mix: direct answering, where the model answers as itself; role-play, where it adopts a sampled persona; and prediction, where it predicts which option that person would pick (Fig.~\ref{fig:inputlevel}d--f). Across DeepSeek-V4-Flash (Fig.~\ref{fig:inputlevel}d), Qwen3-32B (Fig.~\ref{fig:inputlevel}e) and GPT-4.1-mini (Fig.~\ref{fig:inputlevel}f), the ordering direct answering $>$ role-play $>$ prediction holds at the dataset level, category by category, and in the per-question distributions. Two facts follow. First, all three framings stay on the positive side of the human baseline, so the bias is a genuine prior carried across framings, not something specific to persona adoption. Second, third-person prediction lowers the bias more than first-person role-play: asking a model to predict someone else's choice, rather than to be that person, brings out a more detached, less norm-driven stance. Framing is thus a size dial, not a fix.

\header{Aligned models cannot fully play malicious personas} To test whether the bias is a deep prior or just a default voice, we rewrote each sampled persona to encode antisocial, self-interested and harm-tolerant attitudes, and re-ran the role-play protocol for the same three models with matched questions, scoring and decoding (Fig.~\ref{fig:inputlevel}g--i). The three models converge on the same one-sided limit. The malicious rewrite lowers the win rate on fairness optimism, emotional softening and benevolent interpretation, the dimensions less directly targeted by safety training (Fig.~\ref{fig:inputlevel}g,i). But on social desirability, prosociality and harm aversion, the dimensions most shaped by RLHF, it cannot push answers below the human baseline, and sometimes even raises them slightly. For Qwen3-32B, the malicious persona lowers the win rate on all four datasets, yet every dataset stays above 0.5 (Fig.~\ref{fig:inputlevel}h). A persona that should produce a below-average respondent can pull the model toward the human reference, but never past it. On the SDB, PSB and HAB axes, aligned models have measurably lost the ability to imitate people at the less-benevolent end, so any synthetic sample built from them will under-represent that part of the human range, exactly the kind of blind spot a researcher needs to know about and, as we show below, can correct for.

\subsection*{Output level}

\begin{figure*}[!tb]
  \centering
  \includegraphics[width=1\linewidth]{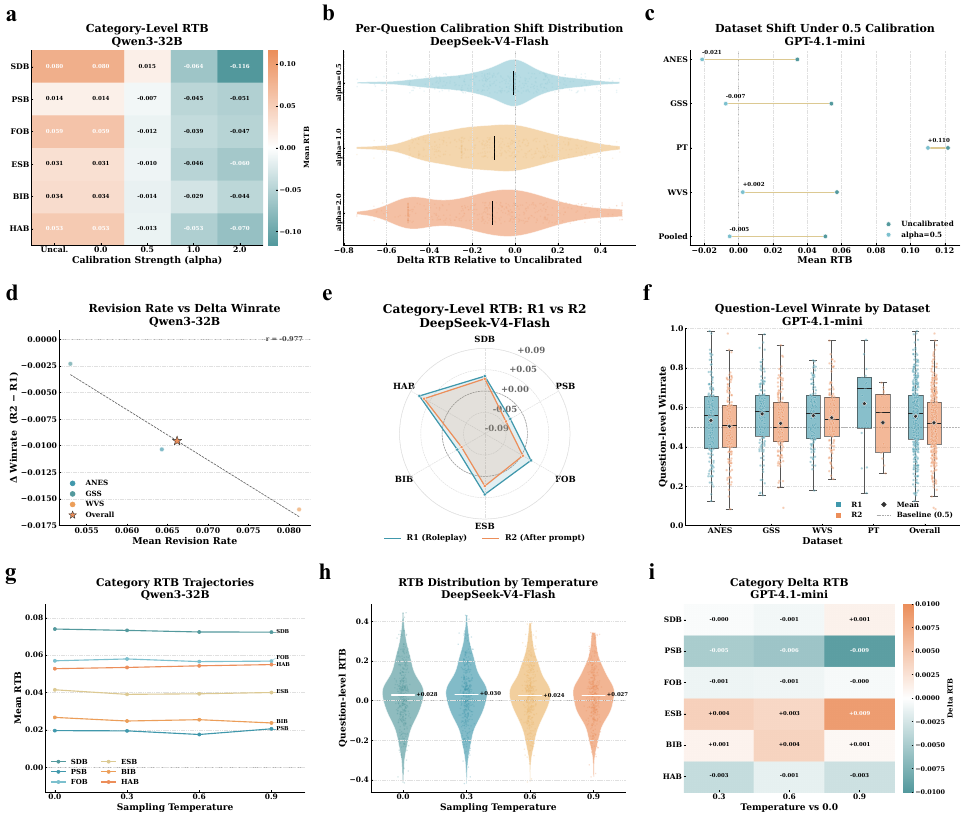}
  \caption{\textbf{Effects of calibration, self-correction and sampling temperature on benevolence bias.}
  \textbf{a}, Category-level BTB of Qwen3-32B across contrastive-calibration strengths ($\alpha$), showing a dose-dependent reduction of the bias on all six dimensions.
  \textbf{b}, Per-question change in BTB of DeepSeek-V4-Flash under contrastive calibration, relative to the uncalibrated baseline.
  \textbf{c}, Mean BTB of GPT-4.1-mini per source dataset at calibration strength 0.5, showing dataset-level differences in calibration effectiveness.
  \textbf{d}, Revision rate and change in win rate of Qwen3-32B under self-correction, decomposed by bias category.
  \textbf{e}, Category-level win rate of DeepSeek-V4-Flash under the baseline (R1) and post-prompt (R2) conditions.
  \textbf{f}, Question-level win rate of GPT-4.1-mini under R1 and R2, split by source dataset, with condition means indicated.
  \textbf{g}, Mean BTB of Qwen3-32B per bias category across sampling temperatures.
  \textbf{h}, Per-question BTB distributions of DeepSeek-V4-Flash across sampling temperatures.
  \textbf{i}, Change in category-level BTB of GPT-4.1-mini as a function of sampling temperature.}
  \label{fig:outputlevel}
\end{figure*}

\header{Contrastive calibration removes the bias, with no retraining} Since the bias survives prompt-level interventions, can it be removed at the output level, without changing the model or the prompt? We developed a contrastive calibration method that works entirely on the next-token probability vector returned for each question, so it is model-agnostic and works with black-box APIs. For each (persona, question) pair we make a second call under a neutral persona, keeping the question and decoding settings fixed; this estimates the model's default answer prior, the option it favours once the persona signal is removed. We then divide the persona-conditioned distribution by this prior raised to the power $\alpha$, and renormalize:
\begin{equation}
P^{*}(a) \;\propto\; \frac{P(a \mid \mathrm{persona}, \mathrm{question})}{P(a \mid \mathrm{neutral}, \mathrm{question})^{\alpha}}.
\label{eq:calibration}
\end{equation}
In plain terms, the method removes the share of the model's preference that it would have shown with no persona at all. The strength $\alpha$ sets the size of the correction: $\alpha = 0$ leaves the answers unchanged, $\alpha = 1$ removes the prior entirely. The neutral persona is a persona-free prompt (``You are a survey respondent.'') with the same format as the persona-conditioned calls (Methods). For Qwen3-32B, sweeping $\alpha$ from 0 to 1 gives a dose-dependent drop on all six dimensions: by $\alpha \approx 0.5$ all six categories collapse into a tight band around zero, and at $\alpha = 1$ several dip below the baseline, showing the correction can overshoot when pushed too hard (Fig.~\ref{fig:outputlevel}a). For DeepSeek-V4-Flash, the per-question $\Delta$BTB shifts left at every $\alpha$, with no question moving the wrong way, and the biggest drops fall on the questions with the largest starting BTB (Fig.~\ref{fig:outputlevel}b). For GPT-4.1-mini at $\alpha = 0.5$, every source dataset moves from clearly positive to near zero without overshooting (Fig.~\ref{fig:outputlevel}c). Contrastive calibration thus reduces the bias across dimensions, questions and datasets, with no retraining, no prompt change and no access to weights, making it a practical, model-agnostic correction whose strength $\alpha$ trades off leftover bias against overshoot.

\header{Aligned models can spot their own bias, but only partly fix it} Can a model recognize its own benevolent shift when challenged? We tested this with a two-round protocol, entirely at the prompt level. In round one, the model answers ($a_1$) under the standard role-play framing; in round two, we reissue the prompt with ``Your answer was $a_1$. Does this truly reflect this person's background, or your own preference? Revise if needed.'', producing $a_2$ (Fig.~\ref{fig:outputlevel}d--f). The revisions are directed, not random. Qwen3-32B preferentially takes back the answers that lean most benevolent, with the biggest drops on social desirability and harm aversion (Fig.~\ref{fig:outputlevel}d). DeepSeek-V4-Flash's win rate falls from R1 to R2 on every dimension but stays above 0.5 on all of them (Fig.~\ref{fig:outputlevel}e). GPT-4.1-mini's distributions shift toward 0.5 on every dataset without crossing it (Fig.~\ref{fig:outputlevel}f). Models thus hold an internal sense of their own bias and act on it when prompted, but this self-awareness neither switches on by itself nor finishes the job: reflection alone cannot restore the human distribution that calibration reaches at $\alpha \approx 0.5$.

\header{The bias does not change with sampling temperature} Finally, we asked whether the bias lives only at the most likely answer or runs through the whole answer distribution, by varying decoding temperature with everything else fixed (Fig.~\ref{fig:outputlevel}g--i). For Qwen3-32B, the category-level mean BTB traces a nearly flat line across the temperature grid, with the category ordering preserved throughout (Fig.~\ref{fig:outputlevel}g). For DeepSeek-V4-Flash, the per-question violins widen a little at higher temperatures but keep the same centre and positive skew (Fig.~\ref{fig:outputlevel}h). For GPT-4.1-mini, the change in mean BTB relative to temperature 0 stays within $\pm 0.01$ in every cell (Fig.~\ref{fig:outputlevel}i). Across the tested range, the direction, size and category profile of the bias are all stable. Benevolence bias therefore sits at the centre of the answer distribution, not in its tails: it vanishes neither when the model picks the mode nor when it explores. This rules out output randomness as a fix, and confirms contrastive calibration as the one intervention tested here that recovers the human reference.

\section*{Discussion}
Across 18 widely used large language models and four major social-science datasets, we find a consistent shift of simulated human answers toward the benevolent end on value-laden questions: 83 of 108 BTB cells and 75 of 108 BWR cells are positive, concentrated on social desirability (mean BWR $=0.565$) and harm aversion (mean BWR $=0.569$), while emotional softening is close to null (mean BWR $=0.494$). Model rankings agree across the three general survey datasets (Spearman's $\rho$ between 0.54 and 0.63), so the benchmark captures a property of the models, not of any one dataset. We call this shift benevolence bias. It is not a thin layer of style on top of an otherwise faithful value representation, but a prior the model brings to every persona it adopts and every survey it answers; naming and measuring it is what makes it something researchers can plan around.

Where the bias does and does not appear points to its source. Temperature moves the category-level BTB by no more than $\pm 0.01$; switching language or framing changes the size of the shift but never its direction; capability explains little of the cross-model variance ($R^2 = 0.16$); and internal thinking softens but does not remove the shift. By elimination, the bias is unlikely to be an artefact of decoding randomness, prompt voice, framing or raw capability. The Qwen2.5 comparison points to the source directly: instruction tuning raised the mean BTB at four of the five sizes tested, with the biggest jump at 32B, and the bias grows with model size in both Qwen3 and Qwen2.5. Benevolence bias thus traces to the post-training stage, and the alignment pipeline installs it more strongly as models grow.

This places benevolence bias inside a larger question: how alignment shapes a model's value representation. A common view is that alignment refines a pre-trained model into more helpful, more acceptable behaviour. Our results suggest that, on value-laden questions, alignment does more than refine: it shifts. The malicious-persona limit makes this concrete. The dimensions less directly targeted by safety training (fairness optimism, emotional softening, benevolent interpretation) can fall below their normal values, so the models retain some ability to produce less benevolent answers there. The dimensions most shaped by RLHF (social desirability, prosociality, harm aversion) never crossed the human baseline, even under explicit instruction. We read this as a narrowing of representational coverage: the model is not merely nudged toward benevolent answers, but has lost some ability to imitate a region of the real range of human attitudes.

This narrowing matters for the fast-growing use of LLMs as silicon samples. Synthetic samples built from current aligned models will over-represent the benevolent end of social-desirability, prosociality and harm-aversion questions, under-represent the other end, and compress the tail variance real surveys show. Because the bias survives the fixes a researcher would try first (persona conditioning, faithful role-play instructions, lower temperature, invitations to reconsider), off-the-shelf use will produce distributions biased not only in the average but in the range; and synthetic data generated for downstream training inherits the same compression, risking a feedback loop in which later models learn from an already shifted value representation. Knowing exactly where this bites is what lets researchers use these tools with their eyes open.

The encouraging part is that the shift can be substantially undone, with no retraining and no access to model weights. At a moderate strength of $\alpha \approx 0.5$, contrastive calibration brings all six dimensions into a tight band around the human baseline, while higher strengths overshoot; the only requirement is a small set of next-token logits, which most current APIs expose. Self-correction, by contrast, is directed but partial: models take back their most benevolent answers when asked to reconsider, yet never return fully to the human reference. The fix that works is distributional, not behavioural, and it is simple enough to drop into existing pipelines today.

These conclusions come with limits worth stating plainly. First, our benchmark draws on datasets that are mostly Western or globally administered (ANES and GSS are US samples; WVS has uneven regional weighting; the prospect-theory replication concerns decision-making), so generalization to value systems outside this frame is not settled. Second, we tested English and Chinese prompts only; how the bias looks in lower-resource languages is open. Third, our questions are multiple-choice with ordered options; whether the same prior shapes free-form answers, where benevolence may surface through wording or topic avoidance, needs a separate design. Fourth, closed-weight alignment recipes are not public, so we can trace the bias to post-training but not to any specific RLHF, DPO or SFT component. Finally, our calibration assumes access to per-token probabilities, which not all APIs expose.

Looking ahead, the central tension is between two goals: alignment optimizes models for socially desirable behaviour, whereas social-science use asks for faithful coverage of the full range of human attitudes, including its uncomfortable parts. Current alignment recipes meet the first goal at a measurable cost to the second, and what post-training compresses away cannot be recovered by prompting. But this is a solvable problem. On the modelling side, alignment objectives that preserve coverage of the human distribution, rather than only shifting its average, could make safety and faithful simulation attainable together. On the measurement side, distributional corrections such as the one proposed here already offer a ready-to-use safeguard and a sharp diagnostic for tracking how alignment reshapes value representation across model generations. With a clear map of the bias and a simple correction in hand, aligned LLMs can move from promising but unproven stand-ins toward dependable instruments for social science.

\section*{Methods}

\subsection*{Benchmark stimulus corpus construction}
We built a psychometrically grounded benchmark by drawing stimuli from established social-science surveys and standardizing them for LLM evaluation. The pipeline has two steps: acquiring the raw data, and decoupling each item from its original context.

\header{Empirical data acquisition and structuring} To keep the stimuli realistic, we drew them from four foundational social-science datasets: the World Values Survey (WVS)\footnote{\url{https://www.worldvaluessurvey.org/}}, the multilingual survey instrument of a prospect-theory replication experiment (PT)\footnote{\url{https://osf.io/vtx3c/}}, the American National Election Studies (ANES)\footnote{\url{https://electionstudies.org/}}, and the General Social Survey (GSS)\footnote{\url{https://gss.norc.org/}}. We write this collection as $\mathcal{D}_{\mathrm{raw}} = \{\mathcal{D}_{\mathrm{WVS}}, \mathcal{D}_{\mathrm{PT}}, \mathcal{D}_{\mathrm{ANES}}, \mathcal{D}_{\mathrm{GSS}}\}$. Because the sources come in many document formats, we use a two-stage extraction pipeline. First, a format-aware parser $\mathcal{P}$ turns each raw document $d$ into plain text: $\mathcal{P}(d) \rightarrow \tilde{d}$. Second, an LLM $\mathcal{M}_{\mathrm{extract}}$ reads the parsed text and produces a structured, machine-readable set of questions. The combined question set $\mathcal{Q}_{\mathrm{raw}}$ is thus
\begin{equation}
\mathcal{Q}_{\mathrm{raw}} \;=\; \bigcup_{d \in \mathcal{D}_{\mathrm{raw}}} \mathcal{M}_{\mathrm{extract}}\!\left(\mathcal{P}(d)\right),
\label{eq:extract}
\end{equation}
where $\mathcal{M}_{\mathrm{extract}}(\cdot)$ maps parsed source text into a single programmatic schema, for example a structured JSON object with the item stem, the response options and metadata. Splitting parsing from semantic extraction keeps the output consistent no matter what format the original source used.

\header{Stimulus standardization and context decoupling} In real surveys, items often use complex structures with shared preambles, such as matrix questions, and human answers are sensitive to option order. To isolate the model's own bias, we use an LLM $\mathcal{M}_{\mathrm{rewrite}}$ to decouple each item $q \in \mathcal{Q}_{\mathrm{raw}}$ from its context and rewrite it as a standalone question. This yields a set of self-contained, independent items $\mathcal{Q}_{\mathrm{indep}}$:
\begin{equation}
\mathcal{Q}_{\mathrm{indep}} \;=\; \left\{ \mathcal{M}_{\mathrm{rewrite}}(q) \;\middle|\; q \in \mathcal{Q}_{\mathrm{raw}} \right\},
\label{eq:rewrite}
\end{equation}
where $\mathcal{M}_{\mathrm{rewrite}}(\cdot)$ turns long, compound survey phrasings into natural, direct, standalone questions. It removes preambles, embedded answer cues and context dependencies, while keeping the original psychometric intent intact. By turning each stimulus into a standalone question that reads like normal conversation, this step ensures the evaluation measures the model's own prior rather than an artefact of a particular survey layout.

\vspace{6pt}
\begin{tcolorbox}[colframe=black!25, colback=black!2, colbacktitle=black!8, coltitle=black,
  fonttitle=\bfseries, arc=2pt, boxrule=0.6pt,
  title={Box 1 $\vert$ Example: context decoupling via an LLM}]
\footnotesize
\headernodot{Original survey item ($q$):}\\
``For each of the following aspects, indicate how important it is in your life. Would you say it is very important, rather important, not very important, or not important at all? --- Friends''

\vspace{6pt}
\headernodot{Rewritten independent item ($q_{\mathrm{indep}}$):}\\
``How important are friends in your life?''

\vspace{6pt}
\headernodot{Preserved response options:}\\
1. Very important \quad 2. Rather important \quad 3. Not very important \quad 4. Not at all important
\end{tcolorbox}
\vspace{6pt}

\subsection*{Taxonomic annotation and cascaded quality validation}
To keep the benchmark psychometrically sound and construct-valid, we define a bias taxonomy $\mathcal{B} = \{b_0, b_1, \ldots, b_k\}$, where $\{b_1, \ldots, b_k\}$ are the $k$ target bias dimensions (the six dimensions of Table~\ref{tab:main}) and $b_0$ is the null category, a neutral stimulus with no target bias signal. Labelling the whole candidate pool by hand is not feasible at scale, both logistically and in terms of rater burden. We therefore use a three-stage cascaded validation pipeline that steadily raises label quality: (1) LLM-based pre-screening, (2) expert heuristic review, and (3) independent researcher consensus.

\header{Stage 1: LLM-based pre-screening} To handle a large volume of candidate items efficiently, we use an LLM $\mathcal{M}_{\mathrm{annotate}}$ as a high-throughput first annotator. For each item $q$, the model assigns a provisional bias label from the taxonomy:
\begin{equation}
b^{(q)}_{\mathrm{LLM}} = \mathcal{M}_{\mathrm{annotate}}(q), \qquad b^{(q)}_{\mathrm{LLM}} \in \mathcal{B}.
\label{eq:annotate}
\end{equation}
Items assigned to the null category ($b^{(q)}_{\mathrm{LLM}} = b_0$) are dropped right away, shrinking the pool to items where at least one bias signal is present. This stage uses the broad semantic coverage of LLMs to make a cheap, scalable first pass.

\header{Stage 2: Expert heuristic review} Because LLMs can still drift semantically or misread context, the surviving items get a quick heuristic review by senior domain experts. Each expert checks whether the provisional label $b^{(q)}_{\mathrm{LLM}}$ is right and whether the item $q$ is coherent. We write this as a binary keep-or-drop indicator:
\begin{equation}
E\!\left(q, b^{(q)}_{\mathrm{LLM}}\right) =
\begin{cases}
1, & \text{if the expert agrees with } b^{(q)}_{\mathrm{LLM}} \text{ and } q \text{ is coherent},\\[2pt]
0, & \text{otherwise}.
\end{cases}
\label{eq:expert}
\end{equation}
Only items with $E = 1$ move to the final stage, which focuses the human effort on the strongest candidates.

\header{Stage 3: Independent researcher consensus} To get high-confidence labels with strict inter-rater reliability, the expert-validated subset goes to $N$ domain-trained doctoral researchers for independent binary verification. Each researcher judges whether item $q$ genuinely and only embodies the assigned bias dimension $b^{(q)}_{\mathrm{LLM}}$. They work fully blinded, with no access to the experts' earlier decisions or to their peers' answers, to avoid authority bias and anchoring. The consensus indicator is
\begin{equation}
C\!\left(q, b^{(q)}_{\mathrm{LLM}}\right) = \operatorname{mode}\left(\left\{ v_i\!\left(q, b^{(q)}_{\mathrm{LLM}}\right) \right\}_{i=1}^{N}\right),
\label{eq:consensus}
\end{equation}
where $v_i \in \{0, 1\}$ is the binary score from researcher $i$. The ground-truth label $\hat{b}_q$ is confirmed as $b^{(q)}_{\mathrm{LLM}}$ only when $C = 1$; items that fail to reach majority consensus are dropped for good.

\header{Benchmark synthesis} The final benchmark $\mathcal{Q}_{\mathrm{final}}$ keeps only the items that pass both the expert gate and researcher consensus:
\begin{equation}
\chi(q) =
\begin{cases}
1, & \text{if } E\!\left(q, b^{(q)}_{\mathrm{LLM}}\right) = 1 \;\wedge\; C\!\left(q, b^{(q)}_{\mathrm{LLM}}\right) = 1,\\[2pt]
0, & \text{otherwise},
\end{cases}
\label{eq:chi}
\end{equation}
\begin{equation}
\mathcal{Q}_{\mathrm{final}} = \left\{ q \in \mathcal{Q}_{\mathrm{indep}} \;\middle|\; \chi(q) = 1 \right\}.
\label{eq:final}
\end{equation}
This double-consensus rule removes items where the algorithm and the human raters disagree, so every retained stimulus captures its intended construct with minimal measurement noise.

\subsection*{Demographic simulation and behavioural distribution estimation}

\header{In-silico demographic simulation} To get a human reference grounded in real population diversity, we use an in-silico demographic conditioning setup. For each dataset $d$, we sample a set of real respondents $\mathcal{U} = \{u_1, \ldots, u_N\}$ and build, for each respondent $u_j$, a demographic profile vector $p_j$ encoding their key socio-economic attributes. This profile then conditions an LLM $\mathcal{M}_{\mathrm{sim}}$ to answer a stimulus $q'$:
\begin{equation}
r_{j,q'} = \mathcal{M}_{\mathrm{sim}}(q', p_j),
\label{eq:sim}
\end{equation}
where $r_{j,q'} \in \{1, \ldots, K\}$ is the predicted option index for respondent $u_j$. Conditioning on $p_j$ steers the model toward the tendencies of the target group rather than its own default prior.

\header{Aggregation into a population-level distribution} We combine the individual simulated answers into a synthetic probability mass function (PMF) over the $K$ options. The estimated mass on option $k$ is
\begin{equation}
F_k = \frac{1}{N} \sum_{j=1}^{N} \mathbb{I}\!\left(r_{j,q'} = k\right).
\label{eq:pmf}
\end{equation}
This gives the population-level synthetic distribution $F$, which serves as the LLM behavioural reference. We compare it directly against the empirical human distribution $G$ from the real respondent sample $\mathcal{U}$, which lets us measure the gap between them quantitatively.

\subsection*{Quantitative metrics for benevolence bias}
We use two complementary metrics that capture benevolence bias from different angles: BTB captures the population-level shift, whereas BWR captures instance-level deviation.

\header{Benevolence Tendency Bias (BTB)} BTB measures the overall directional shift of the LLM answer distribution relative to the human reference. For a stimulus whose $N$ options are ordered by increasing benevolence (option $i = 1$ least benevolent, $i = N$ most benevolent), we assign a position-aware weight
\begin{equation}
W_i = \frac{i}{N},
\label{eq:weight}
\end{equation}
which gives more weight to deviations at the highly benevolent end. Let $F_i$ and $G_i$ be the mass on option $i$ from the LLM and the human reference, respectively. BTB is
\begin{equation}
\mathrm{BTB}(F, G) = \sum_{i=1}^{N} W_i \cdot F_i \;-\; \sum_{i=1}^{N} W_i \cdot G_i.
\label{eq:btb}
\end{equation}
A positive BTB means the model shifts mass toward more benevolent options relative to the human baseline (a positive benevolence bias); a negative BTB means the opposite (a benevolence deficit).

\header{Benevolence Win Rate (BWR)} BTB captures the overall tendency but ignores instance-level variation. BWR complements it by comparing the model's choice against the human answer on each single question. For an evaluation set of $M$ instances, let $y_m^{(k)}$ and $y_g^{(k)}$ be the ordered option indices chosen by the LLM and the human on instance $k$. BWR is
\begin{equation}
\mathrm{BWR} = \frac{1}{M} \sum_{k=1}^{M} \left[ 1 \cdot \mathbb{I}\!\left(y_m^{(k)} > y_g^{(k)}\right) + 0.5 \cdot \mathbb{I}\!\left(y_m^{(k)} = y_g^{(k)}\right) + 0 \cdot \mathbb{I}\!\left(y_m^{(k)} < y_g^{(k)}\right) \right].
\label{eq:bwr}
\end{equation}
The model scores 1 when it picks a strictly more benevolent option than the human, 0.5 for a tie, and 0 otherwise. A BWR above 0.5 means the model is systematically more benevolent than the human on a per-question basis; BWR $=0.5$ means parity; and BWR below 0.5 means the model is consistently less benevolent than the human reference.

\section*{Data availability}
All the used datasets in this study are available on GitHub at \url{xxx}.

\section*{Code availability}
All the source codes to reproduce the results in this study are available on GitHub at \url{xxx}.

\section*{Acknowledgement}
The work is supported by xxxxx.

\section*{Author contributions}
X.X., X.X. and X.X. contributed to the ideation and design of the research; X.X. and X.X. performed the research; all authors contributed to writing and editing the paper.

\section*{Corresponding authors}
Xu Chen (\url{xu.chen@ruc.edu.cn}); Ji-Rong Wen (\url{xxx})

\section*{Competing interests}
The authors declare no competing interests.

\bibliographystyle{unsrt}
\bibliography{references}

@inproceedings{aher2023using,
  title     = {Using large language models to simulate multiple humans and replicate human subject studies},
  author    = {Aher, Gati and Arriaga, Rosa I. and Kalai, Adam Tauman},
  booktitle = {Proceedings of the 40th International Conference on Machine Learning (ICML)},
  year      = {2023}
}

@techreport{anes2020,
  title       = {{ANES} 2020 time series study: Full release},
  author      = {{American National Election Studies}},
  institution = {University of Michigan and Stanford University},
  year        = {2021}
}

@article{argyle2023out,
  title   = {Out of one, many: Using language models to simulate human samples},
  author  = {Argyle, Lisa P. and Busby, Ethan C. and Fulda, Nancy and Gubler, Joshua R. and Rytting, Christopher and Wingate, David},
  journal = {Political Analysis},
  volume  = {31},
  number  = {3},
  pages   = {337--351},
  year    = {2023}
}

@article{atari2023which,
  title   = {Which humans?},
  author  = {Atari, Mohammad and Xue, Mona J. and Park, Peter S. and Blasi, Damian E. and Henrich, Joseph},
  journal = {PsyArXiv preprint},
  year    = {2023},
  doi     = {10.31234/osf.io/5b26t}
}

@article{bai2022training,
  title   = {Training a helpful and harmless assistant with reinforcement learning from human feedback},
  author  = {Bai, Yuntao and Jones, Andy and Ndousse, Kamal and Askell, Amanda and Chen, Anna and DasSarma, Nova and Drain, Dawn and Fort, Stanislav and Ganguli, Deep and Henighan, Tom and others},
  journal = {arXiv preprint arXiv:2204.05862},
  year    = {2022}
}

@article{bai2022constitutional,
  title   = {Constitutional {AI}: Harmlessness from {AI} feedback},
  author  = {Bai, Yuntao and Kadavath, Saurav and Kundu, Sandipan and Askell, Amanda and Kernion, Jackson and Jones, Andy and Chen, Anna and Goldie, Anna and Mirhoseini, Azalia and McKinnon, Cameron and others},
  journal = {arXiv preprint arXiv:2212.08073},
  year    = {2022}
}

@article{bisbee2024synthetic,
  title   = {Synthetic replacements for human survey data? {T}he perils of large language models},
  author  = {Bisbee, James and Clinton, Joshua D. and Dorff, Cassy and Kenkel, Brenton and Larson, Jennifer M.},
  journal = {Political Analysis},
  volume  = {32},
  number  = {4},
  pages   = {401--416},
  year    = {2024}
}

@inproceedings{cao2023assessing,
  title     = {Assessing cross-cultural alignment between {ChatGPT} and human societies: An empirical study},
  author    = {Cao, Yong and Zhou, Li and Lee, Seolhwa and Cabello, Laura and Chen, Min and Hershcovich, Daniel},
  booktitle = {Proceedings of the First Workshop on Cross-Cultural Considerations in NLP (C3NLP) at EACL},
  year      = {2023}
}

@article{casper2023open,
  title   = {Open problems and fundamental limitations of reinforcement learning from human feedback},
  author  = {Casper, Stephen and Davies, Xander and Shi, Claudia and Gilbert, Thomas Krendl and Scheurer, J{\'e}r{\'e}my and Rando, Javier and Freedman, Rachel and Korbak, Tomasz and Lindner, David and Freire, Pedro and others},
  journal = {Transactions on Machine Learning Research; arXiv preprint arXiv:2307.15217},
  year    = {2023}
}

@inproceedings{christiano2017deep,
  title     = {Deep reinforcement learning from human preferences},
  author    = {Christiano, Paul and Leike, Jan and Brown, Tom B. and Martic, Miljan and Legg, Shane and Amodei, Dario},
  booktitle = {Advances in Neural Information Processing Systems (NeurIPS)},
  year      = {2017}
}

@inproceedings{durmus2024towards,
  title     = {Towards measuring the representation of subjective global opinions in language models},
  author    = {Durmus, Esin and Nguyen, Karina and Liao, Thomas I. and Schiefer, Nicholas and Askell, Amanda and Bakhtin, Anton and Chen, Carol and Hatfield-Dodds, Zac and Hernandez, Danny and Joseph, Nicholas and others},
  booktitle = {Conference on Language Modeling (COLM)},
  year      = {2024}
}

@book{edwards1957social,
  title     = {The Social Desirability Variable in Personality Assessment and Research},
  author    = {Edwards, Allen L.},
  publisher = {Dryden Press},
  address   = {New York},
  year      = {1957}
}

@inproceedings{gupta2024bias,
  title     = {Bias runs deep: Implicit reasoning biases in persona-assigned {LLM}s},
  author    = {Gupta, Shashank and Shrivastava, Vaishnavi and Deshpande, Ameet and Kalyan, Ashwin and Clark, Peter and Sabharwal, Ashish and Khot, Tushar},
  booktitle = {International Conference on Learning Representations (ICLR)},
  year      = {2024}
}

@article{hartmann2023political,
  title   = {The political ideology of conversational {AI}: Converging evidence on {ChatGPT}'s pro-environmental, left-libertarian orientation},
  author  = {Hartmann, Jochen and Schwenzow, Jasper and Witte, Maximilian},
  journal = {arXiv preprint arXiv:2301.01768},
  year    = {2023}
}

@article{horton2023large,
  title   = {Large language models as simulated economic agents: What can we learn from homo silicus?},
  author  = {Horton, John J.},
  journal = {NBER Working Paper No. 31122; arXiv preprint arXiv:2301.07543},
  year    = {2023}
}

@book{inglehart2014wvs,
  title     = {World Values Survey: Round Six -- Country-Pooled Datafile},
  author    = {Inglehart, Ronald and Haerpfer, Christian and Moreno, Alejandro and Welzel, Christian and Kizilova, Kseniya and Diez-Medrano, Jaime and Lagos, Marta and Norris, Pippa and Ponarin, Eduard and Puranen, Bi},
  publisher = {JD Systems Institute},
  address   = {Madrid},
  year      = {2014}
}

@article{kirk2024benefits,
  title   = {The benefits, risks and bounds of personalizing the alignment of large language models to individuals},
  author  = {Kirk, Hannah Rose and Vidgen, Bertie and R{\"o}ttger, Paul and Hale, Scott A.},
  journal = {Nature Machine Intelligence},
  volume  = {6},
  pages   = {383--392},
  year    = {2024}
}

@inproceedings{ouyang2022training,
  title     = {Training language models to follow instructions with human feedback},
  author    = {Ouyang, Long and Wu, Jeffrey and Jiang, Xu and Almeida, Diogo and Wainwright, Carroll L. and Mishkin, Pamela and Zhang, Chong and Agarwal, Sandhini and Slama, Katarina and Ray, Alex and others},
  booktitle = {Advances in Neural Information Processing Systems (NeurIPS)},
  year      = {2022}
}

@inproceedings{park2023generative,
  title     = {Generative agents: Interactive simulacra of human behavior},
  author    = {Park, Joon Sung and O'Brien, Joseph C. and Cai, Carrie J. and Morris, Meredith Ringel and Liang, Percy and Bernstein, Michael S.},
  booktitle = {Proceedings of the 36th Annual ACM Symposium on User Interface Software and Technology (UIST)},
  year      = {2023}
}

@inproceedings{perez2023discovering,
  title     = {Discovering language model behaviors with model-written evaluations},
  author    = {Perez, Ethan and Ringer, Sam and Lukosiute, Kamile and Nguyen, Karina and Chen, Edwin and Heiner, Scott and Pettit, Craig and Olsson, Catherine and Kundu, Sandipan and Kadavath, Saurav and others},
  booktitle = {Findings of the Association for Computational Linguistics (ACL)},
  year      = {2023}
}

@article{rafailov2023direct,
  title   = {Direct preference optimization: Your language model is secretly a reward model},
  author  = {Rafailov, Rafael and Sharma, Archit and Mitchell, Eric and Manning, Christopher D. and Ermon, Stefano and Finn, Chelsea},
  journal = {Advances in Neural Information Processing Systems},
  volume  = {36},
  pages   = {53728--53741},
  year    = {2023}
}

@inproceedings{rottger2024political,
  title     = {Political compass or spinning arrow? {T}owards more meaningful evaluations for values and opinions in large language models},
  author    = {R{\"o}ttger, Paul and Hofmann, Valentin and Pyatkin, Valentina and Hinck, Musashi and Kirk, Hannah Rose and Sch{\"u}tze, Hinrich and Hovy, Dirk},
  booktitle = {Proceedings of the 62nd Annual Meeting of the Association for Computational Linguistics (ACL)},
  year      = {2024}
}

@article{rozado2024political,
  title   = {The political preferences of {LLM}s},
  author  = {Rozado, David},
  journal = {PLOS ONE},
  volume  = {19},
  number  = {7},
  pages   = {e0306621},
  year    = {2024}
}

@article{ruggeri2020replicating,
  title   = {Replicating patterns of prospect theory for decision under risk},
  author  = {Ruggeri, Kai and Al{\'i}-Chieh, Sonia and Berkessel, Jana B. and others},
  journal = {Nature Human Behaviour},
  volume  = {4},
  pages   = {622--633},
  year    = {2020}
}

@inproceedings{santurkar2023whose,
  title     = {Whose opinions do language models reflect?},
  author    = {Santurkar, Shibani and Durmus, Esin and Ladhak, Faisal and Lee, Cinoo and Liang, Percy and Hashimoto, Tatsunori},
  booktitle = {Proceedings of the 40th International Conference on Machine Learning (ICML)},
  year      = {2023}
}

@inproceedings{sharma2024towards,
  title     = {Towards understanding sycophancy in language models},
  author    = {Sharma, Mrinank and Tong, Meg and Korbak, Tomasz and Duvenaud, David and Askell, Amanda and Bowman, Samuel R. and Cheng, Newton and Durmus, Esin and Hatfield-Dodds, Zac and Johnston, Scott R. and others},
  booktitle = {International Conference on Learning Representations (ICLR)},
  year      = {2024}
}

@techreport{smith2018general,
  title       = {General social surveys, 1972--2018: Cumulative codebook},
  author      = {Smith, Tom W. and Davern, Michael and Freese, Jeremy and Morgan, Stephen L.},
  institution = {NORC at the University of Chicago},
  year        = {2018}
}

@inproceedings{stiennon2020learning,
  title     = {Learning to summarize with human feedback},
  author    = {Stiennon, Nisan and Ouyang, Long and Wu, Jeff and Ziegler, Daniel and Lowe, Ryan and Voss, Chelsea and Radford, Alec and Amodei, Dario and Christiano, Paul},
  booktitle = {Advances in Neural Information Processing Systems (NeurIPS)},
  year      = {2020}
}

@article{sun2024random,
  title   = {Random silicon sampling: Simulating human sub-population opinion using a large language model based on group-level demographic information},
  author  = {Sun, Seungjong and Lee, Eungu and Nam, Dongyan and Lee, Chanyoung and Park, Sungbin},
  journal = {arXiv preprint arXiv:2402.18144},
  year    = {2024}
}

@inproceedings{wei2022finetuned,
  title     = {Finetuned language models are zero-shot learners},
  author    = {Wei, Jason and Bosma, Maarten and Zhao, Vincent Y. and Guu, Kelvin and Yu, Adams Wei and Lester, Brian and Du, Nan and Dai, Andrew M. and Le, Quoc V.},
  booktitle = {International Conference on Learning Representations (ICLR)},
  year      = {2022}
}

@article{wolf2024fundamental,
  title   = {Fundamental limitations of alignment in large language models},
  author  = {Wolf, Yotam and Wies, Noam and Avnery, Oshri and Levine, Yoav and Shashua, Amnon},
  journal = {Proceedings of the 41st International Conference on Machine Learning (ICML); arXiv preprint arXiv:2304.11082},
  year    = {2024}
}

\clearpage
\setcounter{figure}{0}
\renewcommand{\figurename}{Extended Data Figure}
\setcounter{table}{0}
\renewcommand{\tablename}{Extended Data Table}


\end{document}